\documentclass[onecolumn]{svjour3} 

\usepackage[margin=1in]{geometry} 
\usepackage{color} 
\usepackage{amsmath} 
\usepackage{amsfonts} 
\usepackage{graphicx} 
\usepackage{bm} 
\usepackage[utf8]{inputenc}
\usepackage{authblk} 
\usepackage{multicol} 
\usepackage{booktabs} 
\usepackage{lineno} 
\usepackage{caption} 
\usepackage[numbers,sort&compress]{natbib}

\newlength{\arrow}
\newcommand{\R}{\ensuremath{\mathbb{R}}}           

\renewcommand{\vec}[1]{\bm{\mbox{#1}}}

\newcommand{\PM}{P}

\newcommand{\PD}[1]{\PM_{#1}}

\newcommand{\randvar}[1]{#1}

\newcommand{\randproc}[1]{#1}

\definecolor{red}{rgb}{1,0,0}

\journalname{Journal of the Brazilian Society of Mechanical Sciences and Engineering}

\begin{document}


\title{Robust topology optimization with non-Gaussian material fields using polygonal finite elements}

\titlerunning{}

\author{
Nilton Cuellar \and 
Anderson Pereira \and 
Ivan F. M. Menezes \and 
Americo~Cunha~Jr}

\authorrunning{
N. Cuellar \and 
A. Pereira \and 
I. F. M. Menezes \and 
A. Cunha~Jr}

\institute{N. Cuellar \and A. Pereira \and I. F. M. Menezes \at
				Pontifical Catholic University of Rio de Janeiro -- PUC-Rio,
				Rua Marqu\^{e}s de S\~{a}o Vicente, 225, Rio de Janeiro, 22453-900, RJ, Brazil\\
				\email{nilton.uni@gmail.com}\\
				\email{anderson@puc-rio.br}\\
				\email{ivan@puc-rio.br}
              \and
              A. Cunha~Jr \at
              National Laboratory of Scientific Computing -- LNCC, Av. Getúlio Vargas, 333, Petrópolis, 25651-075, RJ, Brazil \\
              Rio de Janeiro State University -- UERJ,
			  Rua S\~{a}o Francisco Xavier, 524, Rio de Janeiro, 20550-900, RJ, Brazil\\
              \email{americo@lncc.br}\\
}

\date{Received: date / Accepted: date}

\maketitle

\begin{abstract}
We present a computational framework for robust topology optimization that integrates polygonal finite-element discretizations, spatially correlated non-Gaussian material modeling, and non-intrusive polynomial-chaos surrogates. Spatial uncertainty in Young’s modulus is represented as a homogeneous non-Gaussian random field obtained via a memory-less transformation of a truncated Karhunen–Loève expansion, ensuring physical admissibility through positivity of stiffness while preserving the prescribed autocovariance. Polygonal finite elements provide a stable discretization for density-based optimization on unstructured meshes and mitigate checkerboard artefacts and mesh bias, while the sparse polynomial-chaos expansion enables efficient estimation of low-order statistical moments required by the robust objective at a fraction of the cost of intrusive or Monte Carlo approaches. Numerical studies on a cantilever and a curved beam show that introducing non-Gaussian material variability leads to systematic load-path redistribution and a reallocation of 6–12\% of the structural volume, together with a reduction in compliance scatter. The non-intrusive surrogate reproduces intrusive reference results within 3\% using an order of magnitude fewer full finite-element analyses. These results demonstrate that the proposed framework offers a physically consistent and computationally efficient route to topology-optimized designs that remain reliable under realistic material uncertainty.

\end{abstract}


\section{Introduction}
\label{intro}

Topology optimization (TO) seeks the material distribution that renders a structure optimal with respect to a chosen performance metric—most often minimum compliance—while respecting geometric and behavioural constraints \citep{Wu2021p1455,Sigmund2013p1031,Sigmund2004,Eschenauer2001p331}. The ever‐growing demand for lighter, stronger, and more sustainable mechanical components has stimulated a wealth of TO formulations, all of which ultimately require the repeated solution of large, highly non-linear optimization problems. Simultaneously, engineers have recognised that ignoring material and loading variability can produce designs that perform poorly in service. A vigorous body of work, therefore, augments TO with uncertainty quantification (UQ) so that mechanical properties, loads, and manufacturing tolerances are treated as random variables or fields \citep{Doltsinis2004p2221,tootkaboni2012topology,jalalpour2013reliability,keshavarzzadeh2017topology,Gao2022p109238,Li2023p240,Senhora2023p2271,Gresia2024p179,Zheng2024p111990,Luo2024p103583,Guo2025p283}.

Several strategies have been proposed to propagate uncertainty within topology optimization frameworks, each with distinct assumptions and computational implications. Perturbation-based methods rely on truncated Taylor expansions and are computationally attractive, but their accuracy is intrinsically limited to small levels of uncertainty and weak nonlinearities, which restricts their applicability in realistic material-variability settings \citep{smith2014,soize2017}. At the other extreme, Monte Carlo simulation offers great generality and robustness but typically requires an impractically large number of finite-element solves when embedded in an iterative optimization loop \cite{kroese2011,cunhajr2014p1355}. Stochastic collocation and polynomial-chaos-based approaches provide an intermediate alternative, combining spectral accuracy with substantially reduced sampling requirements. In particular, non-intrusive polynomial chaos expansions constructed via sparse-grid quadrature can be interpreted as an efficient form of stochastic collocation, delivering accurate low-order statistics at a fraction of the computational cost of brute-force sampling, while remaining applicable beyond the small-uncertainty regime. These properties make non-intrusive PCE especially well-suited for robust topology optimization under spatially distributed material uncertainty.

Most robust TO studies model uncertainty with \emph{scalar random variables} applied at discrete points. Yet many physical quantities—distributed loads, elastic moduli, plate thicknesses—vary continuously over space and are better idealized as \emph{random fields} \citep{haldar2000reliability,soize2013p2379}. Random fields naturally arise in two- and three-dimensional analyses (plates, shells, solids) as well as in environmental excitations such as wind, waves, and seismic input \citep{grigoriu2012stochastic}.

A widespread simplification is to assume these fields are \emph{Gaussian}, owing to their appealing analytical properties. This assumption is, however, physically inconsistent whenever the quantity of interest is bounded or strictly non-negative. A notable example is Young’s modulus, whose Gaussian modelling permits negative realisations and therefore injects bias into predicted compliance and reliability \citep{soize2017}. While the Karhunen–Loève (KL) expansion discretizes Gaussian fields exactly and economically, it cannot reproduce the heavy or bounded tails that characterise many real materials \citep{lemaitre2010,iaccarino2015}. To overcome that limitation, a memory-less nonlinear mapping can transform a Gaussian KL expansion into a \emph{non-Gaussian random field} with the desired marginal distribution while retaining the target autocovariance \citep{grigoriu2012stochastic}. This strategy preserves the optimality of the KL basis and yields a cost-effective simulator for spatially varying, physically admissible material properties.

On the discretization side, conventional triangular or quadrilateral meshes are prone to checkerboard patterns and mesh bias in density-based TO. \emph{Polygonal finite elements}, with their higher‐order shape-function flexibility and alignment‐free tessellations, mitigate those numerical artefacts and provide a stable platform for gradient-based optimization \citep{Talischi2012p309}. We emphasize that polygonal elements are not the only discretization capable of mitigating checkerboarding or mesh bias, as similar effects can be achieved through alternative interpolation strategies or higher-order standard elements; here, polygonal meshes are adopted as a robust and convenient choice for unstructured tessellations and stable gradient-based topology optimization.

The contribution of this work is the formulation of a robust topology optimization framework in which spatially correlated, non-Gaussian material uncertainty is incorporated in a physically consistent manner within a polygonal finite-element discretization. The framework combines a translation-based non-Gaussian random-field model for Young’s modulus 
\citep{grigoriu2012stochastic,GUILLEMINOT2020385} with topology optimization on unstructured polygonal meshes \citep{antonietti2017,cuelar2018p561,Talischi2009,Talischi2010,talischi2012polytop} and enables efficient uncertainty propagation through a non-intrusive polynomial-chaos surrogate \citep{Eldred2009,keshavarzzadeh2017topology,xiu2002wiener,Xiu2005p1118,xiu2009p242,xiu2010}. This integration is not a straightforward juxtaposition of existing ingredients: it requires a consistent coupling between the stochastic field representation and the polygonal analysis mesh, including a dedicated mapping--interpolation strategy that preserves spatial correlation, statistical consistency, and physical admissibility of the material properties. By resolving these issues within a unified optimization loop, the proposed approach provides a practical and computationally efficient route to robust topology optimization under realistic, non-Gaussian material variability.

Previous studies have combined topology optimization with non-Gaussian material modeling and polynomial-chaos-based uncertainty propagation, but within conventional finite-element discretizations \citep{tootkaboni2012topology,keshavarzzadeh2017topology,Dunning2013p2656}. Conversely, polygonal finite elements have been successfully employed in topology optimization and related uncertainty-aware settings, typically under Gaussian assumptions \citep{cuelar2018p561}. What is missing is a framework that simultaneously combines (i) translation-based non-Gaussian random fields for physically admissible spatial variability and (ii) polygonal finite-element topology optimization on unstructured partitions, including the consistent transfer of field realizations across the two discretizations.

What has been missing—and what we provide here—is a framework in which:
\begin{enumerate}
    \item polygonal finite elements provide a stable discretization for density-based TO on unstructured meshes and mitigate checkerboard artefacts and mesh bias;
    \item a physically consistent, non-Gaussian random field model spatial material variability and obeys positivity of Young’s modulus;
    \item a non-intrusive polynomial‐chaos expansion \citep{Eldred2009,xiu2002p619} accelerates the estimation of low-order statistical moments necessary for robust compliance constraints.
\end{enumerate}

Crucially, combining items (1) and (2) changes the TO answers. In the benchmarks shown later, switching from a Gaussian to a Gamma or Beta modulus field on a polygonal mesh reallocates 6–12 \% of the structural volume and alters the load path in ways that would be suppressed by checkerboard filters on a standard grid. The synergy is therefore threefold: numerical stability, physical fidelity, and computational efficiency.

The remainder of the paper is organised as follows. Section 2 revisits Gaussian random fields, the KL expansion, and the memory-less transformation required for non-Gaussian fields; it also recalls the PC methodology used to estimate statistical measures \citep{Eldred2009,xiu2002p619}. Section 3 outlines deterministic TO, embeds the uncertainty-propagation strategy, derives sensitivity expressions, and states the robust optimization problem. Section 4 presents numerical examples that juxtapose our approach with the methods in \citep{Dunning2013p2656,asadpoure2011robust,zhao2014robust,tootkaboni2012topology,keshavarzzadeh2017topology,cuelar2018p561}. Finally, Section 5 discusses the findings and sketches avenues for further research.


\section{Random fields modeling}
\label{rand_field_mod}

This section presents the fundamental concepts for modeling and computational representation of a random field. Although this material is standard in UQ literature \cite{soize2017,smith2014,grigoriu2012stochastic}, these stochastic tools are still relatively unknown to most engineering, applied science, and numerical optimization communities. Therefore, for the sake of clarity, some important concepts are reproduced in this section.

\subsection{Probabilistic setting}

Consider a probability space $(\Omega, \Im, \PM)$, with sample space $\Omega$, sigma-field $\Im$ and probability measure $\PM$, and let $L_2(\Omega, \PM)$ be the underlying space of the second-order random variables, which is a Hilbert space equipped with the inner product
\begin{equation}
\mathbb{E} \left\lbrace \randproc{X} \randproc{Y} \right\rbrace =
\int \, \int_{\R^2} x \, y \,\, \PD{\randvar{X}, \randvar{Y}} (dx,dy),
\label{01_eq}
\end{equation}
and induced norm
\begin{equation}
\|\randproc{X}\| = \sqrt{\mathbb{E} \left\lbrace \randproc{X}^2 \right\rbrace},
\label{01_eq2}
\end{equation}
where $\mathbb{E} \left\lbrace  \cdot \right\rbrace$ denotes the expectation operator,
and $\PD{\randvar{X}, \randvar{Y}} (dx,dy)$ is the joint distribution
of the random variables $\randvar{X}, \randvar{Y} \in L_2 (\Omega,\PM)$.

\subsection{Random field characterization}
\label{ran_field}

In this probabilistic setting we can define  a $\R^d$-valued random field $\randproc{H}(\vec{x},\omega ):  \mathcal{D} \times \Omega \mapsto \R^d$, that can be thought of as a random vector in $\R^d$, indexed by a spatial coordinate $\vec{x} \in \mathcal{D} \subset \R^n$, $n \geq 1$, defined for every elementary event $\omega \in \Omega$.

If this random field is Gaussian, it is completely characterized by the second-order statistics, i.e., the expectation function $\mu_{\randproc{H}}(\vec{x})$, and the autocovariance function $C_{\randproc{H}\randproc{H}}(\vec{x}_1,\vec{x}_2)$, which generates, for $\mu_{\randproc{H}}(\vec{x}_1) = \mu_{\randproc{H}}(\vec{x}_2) = 0$, the autocorrelation function
\begin{equation}
R_{\randproc{H}\randproc{H}}(\vec{x}_1,\vec{x}_2) = \mathbb{E}\left[ \randproc{H}(\vec{x}_1,\omega) \randproc{H}(\vec{x}_2,\omega) \right].
\label{05_eq}
\end{equation}

On the other hand, if the random field is non-Gaussian, its characterization is more challenging, since it is not uniquely determined by the first two statistical moments \citep{Stefanou2009p1031}.

The random field $\randproc{H}$ is called homogeneous (in the wide sense) if its expectation function is invariant for changes in the space coordinate $\vec{x}$, and the correlation function depends only on the distance between $\vec{x}_1$ and $\vec{x}_2$, i.e.,
\begin{equation}
R_{\randproc{H}\randproc{H}}(\vec{x}_1,\vec{x}_2) = R_{\randproc{H}\randproc{H}}(\mid\vec{x}_1-\vec{x}_2\mid),
\label{08_eq}
\end{equation}
a characteristic that make this type of random field exhibits a certain symmetry in the spatial domain.

\subsection{Gaussian random field generation}

For computational generation purposes, a Gaussian random field must be represented (discretized) in terms of a finite number of random variables. This can be is done with the aid of the KL expansion \citep{ghanem2003, xiu2010}
\begin{equation}
\randproc{H}\left(\vec{x},\omega \right) = 
\mu_H(\vec{x}) + \sum_{i=1}^{\infty} \sqrt{\lambda_i} \, \phi_i(\vec{x}) \, \xi_i(\omega),
\label{09_eq}
\end{equation}
where $\lambda_i$ and $\phi_i(\vec{x})$ respectively denote 
the eigenvalues and eigenfunctions of the autocovariance function, i.e.,
\begin{equation}
\int_{\mathcal{D}} C_{HH}(\vec{x}_1,\vec{x}_2)\phi_i(\vec{x}_1) d\vec{x}_1 = \lambda_i\phi_i(\vec{x}_2),
\label{11_eq}
\end{equation}
respecting the orthonormality condition
\begin{equation}
\int_{\mathcal{D}} \phi_i(\vec{x}) \phi_j(\vec{x}) d\vec{x} = \delta_{ij},
\label{12_eq}
\end{equation}
and $\left\lbrace \xi_i \right\rbrace_{i=1}^{\infty}$ is a set of centered, mutually uncorrelated random variables, i.e.,
\begin{equation}
\mathbb{E} \left\lbrace \xi_i \right\rbrace = 0,
~~\mbox{and}~~
\mathbb{E} \left\lbrace \xi_i \, \xi_j \right\rbrace = \delta_{ij},
\label{13_eq}
\end{equation}
\noindent
where $\delta_{ij}$ is the Kronecker delta function.

In practical terms, a truncation of the series involving $N$ random variables is considered
\begin{equation}
\randproc{H} (\vec{x},\omega) \approx 
\mu_{H}(\vec{x}) + \sum_{i=1}^N \sqrt{\lambda_i} \, \phi_i(\vec{x}) \, \xi_i(\omega),
\label{15_eq}
\end{equation}
with the accuracy being guaranteed by the property which ensures that the mean square truncation error decreases monotonically with the increase in the number of terms of this expansion. 

\begin{figure*}[!htb]
	\centering
	\includegraphics[scale=0.5]{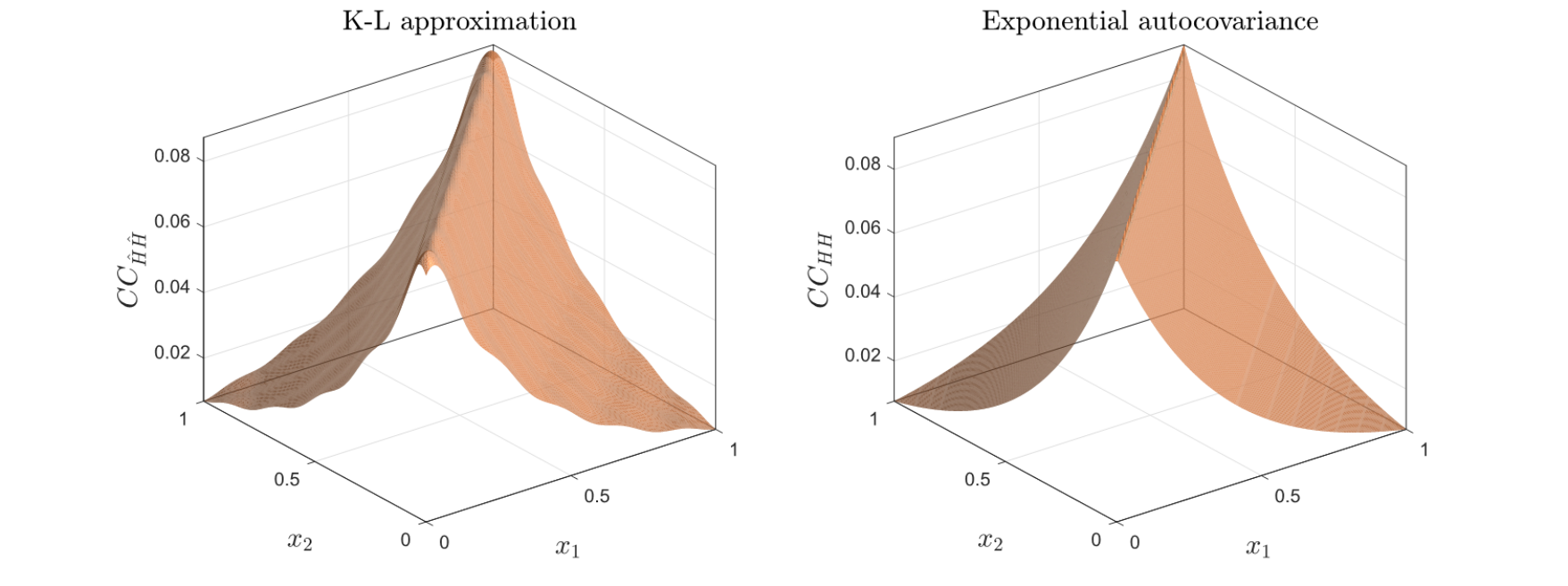}
	\caption{Comparison between an exact exponential autocovariance function (right) and an approximation obtained with 8-term KL expansion (left).}
	\label{fig01}
\end{figure*}

The selection of $N$  depends on the shape of the autocovariance function. More correlated is the random field, smaller the $N$ required to achieve a good accuracy.  Conversely, if the random field is less correlated, a higher $N$ is necessary. For the sake of illustration, Figure~\ref{fig01} shows the graph of an exponential autocovariance function (right) and its approximation using a KL expansion with $N=8$ terms (left). Note that, even with a reasonably coarse discretization, KL expansion produces an accurate representation.

\subsection{Non-Gaussian random field generation}

The computational strategy employed in this work to generate homogeneous non-Gaussian random fields uses a nonlinear transformation (translation process) of a homogeneous Gaussian field \citep{grigoriu2012stochastic}.

The translation of the homogeneous, zero-mean and unit-variance random field $\randproc{H}(\vec{x},\omega)$ is defined in terms of the (memoryless) translation map\footnote{This is a real-valued differentiable, increasing monotonically, nonlinear transformation.} $g(\cdot)$, that gives rise to a homogeneous non-Gaussian random field of the form 
\begin{equation}
\randproc{Y}(\vec{x},\omega) = g \left( \randproc{H}(\vec{x},\omega) \right).
\end{equation}

This nonlinear map may be chosen as $g = F^{-1}_Y \circ \Phi$, where $F^{-1}_Y(\cdot)$ denotes the inverse of a prescribed marginal cumulative distribution function (CDF) for the non-Gaussian field $\randproc{Y} (\vec{x},\omega)$, and $\Phi$ is the CDF of the standard Gaussian distribution, in a way that the homogeneous non-Gaussian random field is given by
\begin{equation}
\randproc{Y}(\vec{x},\omega) = F^{-1}_Y \left( \Phi \left(\randproc{H}(\vec{x},\omega)\right) \right).
\label{17_eq}
\end{equation}


\section{Polynomial chaos surrogate}
\label{pol_chaos}
Polynomial chaos expansion (PCE) is a powerful method to construct computational representations of a function of random variables in terms of a sum of multidimensional orthogonal polynomials of independent random variables weighted by deterministic coefficients that need to be determined \citep{ghanem2003,xiu2002p619,xiu2010,Ghanem2017chap}.

According to PCE theory, a general second-order random function $\randproc{X}(\bm{\xi}) \in L_2(\Omega,P)$, which is parametrized by a set independent random variables $\bm{\xi} = \{\xi_i(\omega)\}_{i=1}^\infty$, is represented by the following expansion
\begin{equation}
\randproc{X}(\bm{\xi}) = \sum^\infty_{j=0}\hat{u}_j \Psi_j(\bm{\xi}),
\label{24_eq}
\end{equation}
\noindent
where $\hat{u}_j$ are the deterministic coefficients to be determined and $\Psi_j(\bm{\xi})$ are the multidimensional polynomials in the random variables $\bm{\xi}$, which constitute a complete orthonomal basis in $L_2(\Omega,P)$.

Note that the series expansion exhibited in Eq.(\ref{24_eq}) involves an infinite collection of $\xi_i$  and $\Psi_j$. For computational practice, it is necessary restrict the representation to a finite number of random variables $\bm{\xi} = \{\xi_i(\omega)\}_{i=1}^N$ and fix the polynominal order to a certain natural $p$. Thus, the PCE with dimension $N$ and order $p$ is truncated with a finite number $(M+1)$ of polynomial terms, where
\begin{equation}
M+1 = {{N+p}\choose{p}} = \frac{(N+p)!}{N! \, p!},
\label{25_eq}
\end{equation}
so that the truncated expression of $\randproc{X}$ is written as
\begin{equation}
\randproc{X}(\bm{\xi}) \approx \sum^M_{j=0}\hat{u}_j \Psi_j(\bm{\xi}),
\label{26_eq}
\end{equation}
where the coefficients are analytically expressed in terms of an orthogonal projection
\begin{equation}
\hat{u}_j = \mathbb{E}\left[\randproc{X}(\bm{\xi})\Psi_j(\bm{\xi}) \right],
\label{27_eq}
\end{equation}
so that they can be calculated using non-intrusive techniques for numerical integration like variants of the Monte Carlo method, tensor product based or ``sparse grid'' quadrature formulas \citep{xiu2009p242,xiu2010,iaccarino2015,Ghanem2017chap}.

The tensor product based quadrature schemes may achieve an exponential convergence in low stochastic dimension, but for a large number of random variables (large may be bigger than 3), a typical scenario when dealing with random fields discretization, the collocation points number grows very rapidly and the rate of convergence deteriorates drastically, a phenomenon known as ``curse of dimensionality'' that makes the computational cost-prohibitive. A possible alternative to the last approach is the use of ``sparse grids'' based quadrature formulas, like the classical Smolyak scheme, that integrates multidimensional functions using a nested quadrature, which defines a reduced grid of points constructed recursively, from an univariate Gaussian quadrature method \citep{gerstner1998numerical,Xiu2005p1118}. Other possibility is to use Monte Carlo method \cite{kroese2011,cunhajr2014p1355} (or its variants) to obtain the PCE coefficients through a regression process \citep{xiu2010,Ghanem2017chap}. This paper to computes the polynomial chaos coefficients using the last two above-mentioned approaches, sparse grids for the cases with low stochastic dimension, and Monte Carlo method when the problem is high-dimensional. 


\section{Topology optimization framework}
\label{top_opt_frame}

\subsection{Deterministic topology optimization }

Topology optimization seeks to determine an optimal distribution of material layout, within a specified design domain, subject to design specifications (e.g. volume or compliance bounds) \citep{Sigmund2004}. In this paper, the objective of the optimization is to minimize the volume of the structure respecting an upper limit for the system compliance. 
The governing partial differential equations (PDEs) under the assumption of static and linear-elastic behavior are solved using the finite element method, resulting in the following linear system of equations
\begin{equation}
\vec{K}(\bm{\rho})\vec{u} = \vec{f}
\end{equation}
where $\vec{u}$  and $\vec{f}$  are the global displacement and loading vectors, respectively; $\vec{K}(\bm{\rho})$ represents the global stiffness matrix, which is dependent on the design variable $\bm{\rho}$ (density). 

To use gradient-based optimization algorithms, we employ a topology optimization formulation known as Solid Isotropic Material with Penalization (SIMP), which defines the stiffness of the intermediate densities of a given element $e$ employing a penalization, i.e.,
\begin{equation}
E(\rho_{e})=\left[ E_{\min}+\left(1-E_{\min}\right){\rho_{e}}^p\right] E^0,
\label{33_eq}
\end{equation}
\noindent
where $0\leq E_{\min}\ll 1$ defines a compliant material used to fill the void regions to enforce existence of solutions to the boundary value problem; $p \geq 1$ is the penalization factor with a value larger than one and ${E}^0$ is the original Young's modulus of the material.

Hence, the stiffness matrix can be assembled as
\begin{equation}
\vec{K}(\bm{\rho}) = \sum_{e=1}^{Ne} E(\rho_{e})\vec{k}^0_{e}
\label{kglobasse}
\end{equation}
where $\vec{k}^0_{e}$ is the element stiffness matrix with unit modulus of elasticity and $N$ is total number of elements in the finite element mesh. 

In this way, the deterministic (baseline) topology optimization problem is formulated as
\begin{equation}
\begin{aligned}
&\!\min_{\bm{z}\;\in \;\left[0,1\right]^Ne}        &\qquad& V(\bm{z}) = 
\frac{1}{|\Omega|}\sum_{e=1}^{Ne}\rho_e v_e \\
&\text{subject to} &      & C(\bm{z}) = \vec{f}^{\,\,T} \, \vec{u} (\bm{z})\leq C_S,\\
& \text{with}                 &      & \vec{K}(\bm{\rho})\vec{u} = \vec{f},\\
\end{aligned}
\label{31_eq}
\end{equation}
where $V(\bm{z})$  is the volume of the design domain to be minimized; $C(\bm{z})$  is the compliance of the structure; and the parameter $C_s$ is the specified highest allowable compliance. The discrete density values, $\rho_e$, are defined in a discrete form via a regularization filter, as follows
\begin{equation}
\bm{\rho}(\bm{z}) = \bm{P}\bm{z}
\label{filtering}
\end{equation}
\noindent
where
\begin{equation}
P_{ij}=\frac{w_{ij}}{\sum_{k=1}^{Ne}w_{ik}v_k}
\end{equation}
is the filter matrix, defined as
\begin{equation}
w_{ij}=\max\left(1-\frac{\|\vec{x}_i-\vec{x}_j\|_2}{R},0\right)
\end{equation}
where $R$ is the filter radius and $\|\vec{x}_i-\vec{x}_j\|_2$ is the Euclidean distance between the centroids, $\vec{x}_i$ and $\vec{x}_j$, of elements $i$ and $j$, respectively.

We note that the linear density filter in Eq.~(\ref{filtering}) is introduced as a regularization operator in the standard setting of density-based topology optimization. Its role is to suppress mesh-scale oscillations in the design field, ensure a well-posed optimization problem, and provide stable sensitivity information for gradient-based updates. This regularization becomes particularly important in the present robust setting, where uncertainty propagation is embedded in the optimization loop and the objective and constraints depend on statistical measures estimated from the stochastic surrogate. Consequently, the present formulation operates in a relaxed (continuous) density parametrization, and intermediate density values may appear as part of the converged solution. The goal of this work is the robust assessment of load paths, volume allocation, and compliance statistics under spatially correlated non-Gaussian material uncertainty, rather than the enforcement of strictly binary (0--1) designs. Enforcing strict discreteness would require additional mechanisms beyond the linear filter (e.g., projection/continuation strategies or perimeter-type regularization), which are outside the scope of the present study.

\subsection{Robust topology optimization}
\label{rob_top_opt}

Once the variability in the material properties can significantly affect the optimal design, taking into account the material uncertainties is essential in any robust design framework \citep{soize2013p2379,soize2017,cuelar2018p561,GUILLEMINOT2020385}. Several definitions of robustness have been proposed in the literature \citep{beyer2007p3190,Birge2011,Doltsinis2004p2221,Shin2011p248}, and the weighted sum of both the mean and standard deviation of the objective function and constraints are generally considered. Then, the trade-off between those two statistical measures could describe the final design, i.e., a design that is less conservative and with a significantly smaller range of variation. Note that it is possible to combine these two statistics into the objective function to improve the structural performance and the design by minimizing the influence of variability owing to the uncertainties. 

In this framework for robust topology optimization (RTO), the underlying material uncertainties are taken into account through a parametric probabilistic approach \cite{soize2013p2379,soize2017}, so that the compliance constraint becomes stochastic. This is not the case of the objective function from Eq.(\ref{31_eq}), since the volume does not depend on the material properties related to stiffness. Therefore, the robust optimization problem is formulated as
\begin{equation}
\begin{aligned}
&\!\min_{\bm{z}\;\in \;\left[0,1\right]^Ne}        &\qquad& V(\bm{z}) = 
\frac{1}{|\Omega|}\sum_{e=1}^{Ne}\rho_e v_e \\
&\text{subject to} &      & \tilde{C} = \mu_{C} (\bm{\rho},\vec{x}) + k \, \sigma_{C} (\bm{\rho},\vec{x})  \leq C_s,\\
& \text{with}                 &      & \vec{K}(\bm{\rho},\vec{x})\vec{u}(\bm{\rho},\vec{x}) = \vec{f},\\
\end{aligned}
\label{34_eq}
\end{equation}
\noindent
where $\vec{x}$ is the spatial coordinate of random field; $\tilde{C}$  is the objective function of the RTO problem to be minimized; $k\geq 0$  is a weighting factor for the objective function; $\mu_{C}$  and $\sigma_{C}$ represent the mean and standard deviation of compliance, respectively; and $C = C(\bm{\rho},\vec{x},\bm{\xi})$  is the stochastic compliance according to the uncertainties in the material properties.

The solution of this RTO problem is computed with the aid of a polynomial chaos surrogate, constructed in a non-intrusive way, such as described in section~\ref{pol_chaos}. This approach is an efficient tool owing to its rapid converge property and its capability to propagate uncertain parameters in the form of random variables and fields.

\subsection{Modeling material uncertainties}

To take into account its spatial variability, Young’s modulus is considered a 2D non-Gaussian univariate multidimensional random field, which according to Eq.(\ref{17_eq}) is represented by
\begin{equation}
\vec{E}^0(\vec{x},\omega)=F^{-1}_E\circ\Phi\left(H(\vec{x},\omega)\right).
\label{36_eq}
\end{equation}

Typically, $F^{-1}_E$ can involve a log-normal, gamma, or uniform distribution for modeling the uncertainties in material properties. In this work, the uniform, generalized beta, and gamma distributions are used for being the simplest among the marginal distributions that are physically consistent. The use of other types of marginal distributions is straightforward.

\begin{figure*}[h]
	\centering
	\includegraphics[scale=0.65]{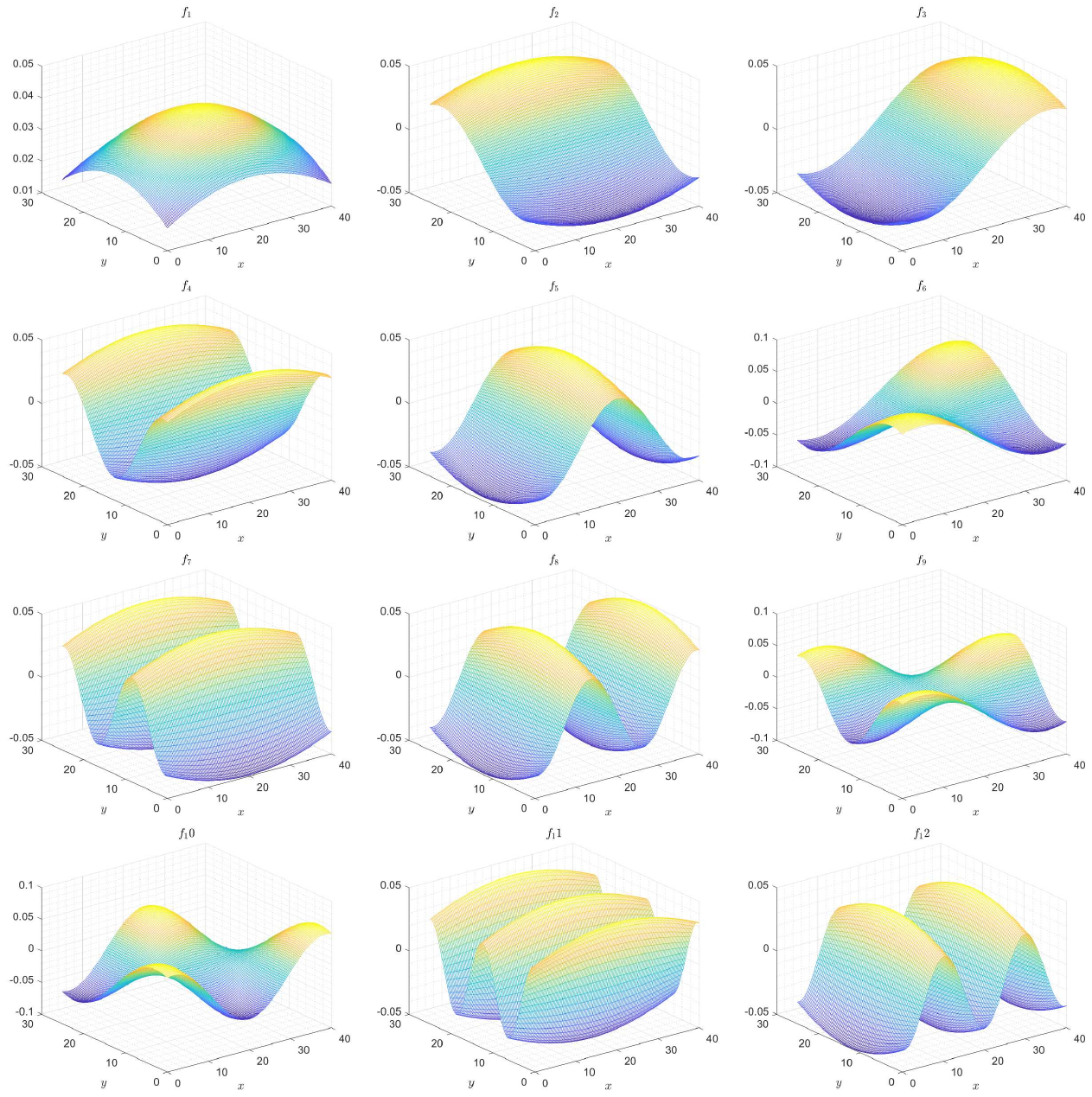}
	\caption{Illustration of the first 12 eigenfunctions of exponential autocovariance function.}
	\label{fig02}
\end{figure*}

To discretize a non-Gaussian field, the method of translation of field is used and can be interpreted as a memoryless nonlinear transformation of Gaussian field $H(\vec{x},\omega)$, which is defined by the 2D Gaussian autocovariance kernel
\begin{equation}
C_{HH}(\vec{x}_1,\vec{x}_2) = \exp\left(-\frac{\left(x_1-y_1\right)^2}{l^2_1}-\frac{\left(x_2-y_2\right)^2}{l^2_2}\right),
\label{39_eq}
\end{equation}
\noindent
where $l_1$ and $l_2$ are the correlation lengths parameters in different directions. An illustration of the first 12 eigenfunctions of this autocovariance function can be seen in the Figure~\ref{fig02}.

\begin{figure*}[!htb]
	\centering
	\includegraphics[scale=2]{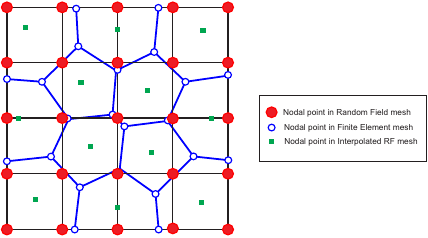}
	\caption{Illustration of RF and FE nodal points used for the mapping interpolation.}
	\label{fig03}
\end{figure*}

To obtain reasonable accuracy, two types of meshes are employed here, one for the random field and the other for finite element discretization, albeit with an identical boundary shape. A structured quadrilateral mesh is used for the random field, with the field values associated with the nodal points of the mesh. For solving the topology optimization problem, polygonal finite element meshes are used because of the well-known advantages of this type of discretization, such as suppressing checkerboard patterns and reducing mesh dependency \cite{Talischi2009,Talischi2010,talischi2012polytop,antonietti2017}. As the random field mesh is different from the FE mesh, a mapping-interpolation method is required to transfer the spatially-distributed material properties to the centroid of the FE mesh. The fundamental concept underlying the mapping-interpolation method is shown in Figure~\ref{fig03}.

More precisely, the random field discretization and the finite element discretization play distinct and complementary roles. The structured quadrilateral mesh is introduced exclusively to represent the Gaussian random field $H(\vec{x},\omega)$ through the truncated Karhunen--Lo\`eve expansion and to apply the translation map $F_E^{-1}\circ\Phi(\cdot)$, yielding nodal samples of the non-Gaussian Young's modulus field $\vec{E}^0(\vec{x},\omega)$ on that structured grid. In contrast, the topology optimization problem is solved on an unstructured polygonal finite element mesh, on which the stiffness matrix, compliance, and sensitivity quantities are assembled.

Since these two discretizations do not coincide, a consistent transfer of the spatially varying material property is required. For each polygonal element $e$, a unique value of Young's modulus $E^0_e(\omega)$ must be assigned to define the element constitutive response and the SIMP penalization. This value is obtained by evaluating the translated random field at the centroid $\vec{x}_e^{\,c}$ of the polygonal element through interpolation on the quadrilateral random-field mesh. Let $\{\vec{x}_a\}_{a=1}^{4}$ denote the four vertices of the quadrilateral cell containing  $\vec{x}_e^{\,c}$, and let $\{E^0(\vec{x}_a,\omega)\}_{a=1}^{4}$ be the corresponding nodal field values. The element-wise modulus is then approximated as
\begin{equation}
E^0_e(\omega) \;=\; E^0(\vec{x}_e^{\,c},\omega) 
\;\approx\; \sum_{a=1}^{4} N_a(\xi,\eta)\,E^0(\vec{x}_a,\omega),
\end{equation}
where $N_a(\xi,\eta)$ are the standard bilinear shape functions of the parent quadrilateral, evaluated at the local coordinates $(\xi,\eta)$ associated with the centroid $\vec{x}_e^{\,c}$. This centroid-based bilinear interpolation ensures a smooth and physically consistent transfer of the non-Gaussian spatial variability from the structured random-field mesh to the polygonal finite element mesh, without altering the Karhunen--Lo\`eve basis, the translation process, or the polygonal discretization 
employed in the topology optimization.

\subsection{Low order statistics estimators}

The calculation of the low-order statistics (mean and standard deviation) of the compliance function $C(\bm{\rho},\vec{x},\bm{\xi})$ can be done efficiently through PCE from Eq.(\ref{26_eq}), that explores the orthonormality of this expansion to build estimators that are easily calculated using the surrogate coefficients. 

In practical terms, an estimator for the mean $\mu_{C}$ is given by
\begin{equation}
\hat{\mu}_{C} (\bm{\rho},\vec{x}) =  \hat{u}_0 (\bm{\rho},\vec{x}),
\label{43_eq}
\end{equation}
while the estimator for the variance $\sigma^2_{C}$ read as 
\begin{equation}
\hat{\sigma}^2_{C} (\bm{\rho},\vec{x}) = \sum^M_{j=0}\hat{u}^2_j (\bm{\rho},\vec{x}) - \hat{u}^2_0 (\bm{\rho},\vec{x}).
\label{45_eq}
\end{equation}

\subsection{Sensitivity analysis}

The use of gradient-based methods to solve Eq.(\ref{34_eq}) requires the computation of gradients of both objective and constraint functions with respect to the design variables $\textbf{z}$. Following the same procedure of \cite{talischi2012polytop}, we can compute the sensitivity of the objective function using the chain rule. We first compute the gradient of the objective function with respect to the density variable ${\rho}_e$, which is expressed as
\begin{equation}
 \frac{\partial V}{\partial \rho_e} = \frac{v_e}{|\Omega|}.
 \label{46_eq}
\end{equation}

By further incorporating the density filtering operation, Eq.(\ref{filtering}), we arrive at the expression for the gradient of the objective function with respect to the design variable $\textbf{z}$ as
\begin{equation}
\frac{\partial V}{\partial \textbf{z}}=\mathbf{P}^T\frac{\partial V}{\partial \boldsymbol{\rho}}.
\end{equation}

The gradient of the compliance constraint function with respect to the design variable $\textbf{z}$ is given as
\begin{equation}
\frac{\partial C}{\partial \textbf{z}}=\mathbf{P}^T\frac{\partial C}{\partial \boldsymbol{\rho}},
\end{equation}
where ${\partial C}/{\partial \boldsymbol{\rho}}$ can be obtained using the adjoint method \citep{Sigmund2004,Talischi2012p309}. Starting from the mechanical equilibrium equation
\begin{equation}
\vec{K}(\bm{\rho},\vec{x})\vec{u}(\bm{\rho},\vec{x}) = \vec{f},
\label{47_eq}
\end{equation}
the compliance $C$ can be represented by
\begin{equation}
C=\vec{f}^T\vec{u}(\bm{\rho},\vec{x})-\bm{\lambda}^T\left(\vec{K}(\bm{\rho},\vec{x})\vec{u}(\bm{\rho},\vec{x})-\vec{f}\right),
\label{47a_eq}
\end{equation}
\noindent
where $\bm{\lambda}$ is an arbitrary vector.  Taking the derivatives of Eq.(\ref{47a_eq}) with respect to the design variable, the following expression is obtained
\begin{equation}
\frac{\partial C}{\partial\rho_e} = \left(\vec{f}^T-\bm{\lambda}^T\vec{K} \right)\frac{\partial\vec{u}}{\partial\rho_e}-\bm{\lambda}^T\frac{\partial\vec{K}}{\partial\rho_e}\vec{u}.
\label{48_eq}
\end{equation}

To avoid having to calculate ${\partial \vec{u}}/{\partial\rho_e}$, we construct an adjoint problem that solves for the adjoint variable

\begin{equation}
\vec{f}^T-\bm{\lambda}^T\vec{K}=\vec{0}.\,
\label{49_eq}
\end{equation}
where one can directly obtain that $\bm{\lambda} = \vec{u}$  (the minimum compliance is self-adjoint), so the derivative of compliance concerning the element design variable given by
\begin{equation}
\frac{\partial C}{\partial\rho_e} = -\vec{u}^T\frac{\partial\vec{K}}{\partial\rho_e}\vec{u}.
\label{50_eq}
\end{equation}

Once the global stiffness matrix from Eq.(\ref{kglobasse}) can be expressed as
\begin{equation}
\frac{ \partial\vec{K}(\bm{\rho}) }
     { \partial\rho_j} = \sum_{e=1}^{N}\frac{\partial E(\rho_{e})}{\partial\rho_j}\vec{k}^0_{e} =
     \frac{\partial E(\rho_{e})}{\partial\rho_e}\vec{k}^0_{e},
\label{50_1_eq}
\end{equation}
one can replace Eq.(\ref{50_1_eq}) into the Eq.(\ref{50_eq}), so that we have
\begin{equation}
\frac{\partial C}{\partial\rho_e} =-\vec{u}_e^T\frac{\partial E(\rho_{e})}{\partial\rho_e}\vec{k}^0_{e}\vec{u}_e.
\label{50_2_eq}
\end{equation}  

Differentiating Eq.(\ref{33_eq}) with respect to $\rho_{e}$, we have
\begin{equation}
\frac{\partial E(\rho_{e})}{\partial\rho_e}=
p\left(1-E_{\min}\right){\rho_{e}}^{p-1} E^0,
\label{50_3eq}
\end{equation}
in way that, by substituting Eq.(\ref{50_3eq}) in Eq.(\ref{50_2_eq}), we obtain the general formulation for the compliance sensitivity, which is given by
\begin{equation}
\frac{\partial C}{\partial\rho_e} = -p\left(E_0 - E_{\min} \right)\rho^{p-1}_e\vec{u}^T_e\vec{k}^0_e\vec{u}_e.
\label{51_eq}
\end{equation}

Thus, the compliance constraint sensitivity with respect to the design variable $\rho_e$ is expressed as
\begin{equation}
\frac{\partial \tilde{C}}{\partial\rho_e} = \frac{\partial \mu_{C}}{\partial\rho_e}(\bm{\rho},\vec{x}) + k \, \frac{\partial \sigma_{C}}{\partial\rho_e} (\bm{\rho},\vec{x}),
\label{52_eq}
\end{equation}
so that efficient statistical estimators for the two terms on the right side can be constructed from the coefficients of the PCE surrogate, i.e.,  
\begin{equation}
\frac{\partial \hat{\mu}_{C}}{\partial\rho_e} (\bm{\rho},\vec{x}) = 
\frac{\partial \hat{u}_0}{\partial\rho_e} (\bm{\rho},\vec{x}),
\label{53_eq}
\end{equation}
and
\begin{equation}
\frac{\partial \hat{\sigma}^2_{C}}{\partial\rho_e} (\bm{\rho},\vec{x}) =
\sum^M_{j=0}\frac{\partial \hat{u}^2}{\partial\rho_e} (\bm{\rho},\vec{x}) - \frac{\partial \hat{u}^2_0}{\partial\rho_e} (\bm{\rho},\vec{x}).
\label{54_eq}
\end{equation}

\subsection{Computational implementation}

\begin{figure*}[!htb]
	\centering
	\includegraphics[scale=0.8]{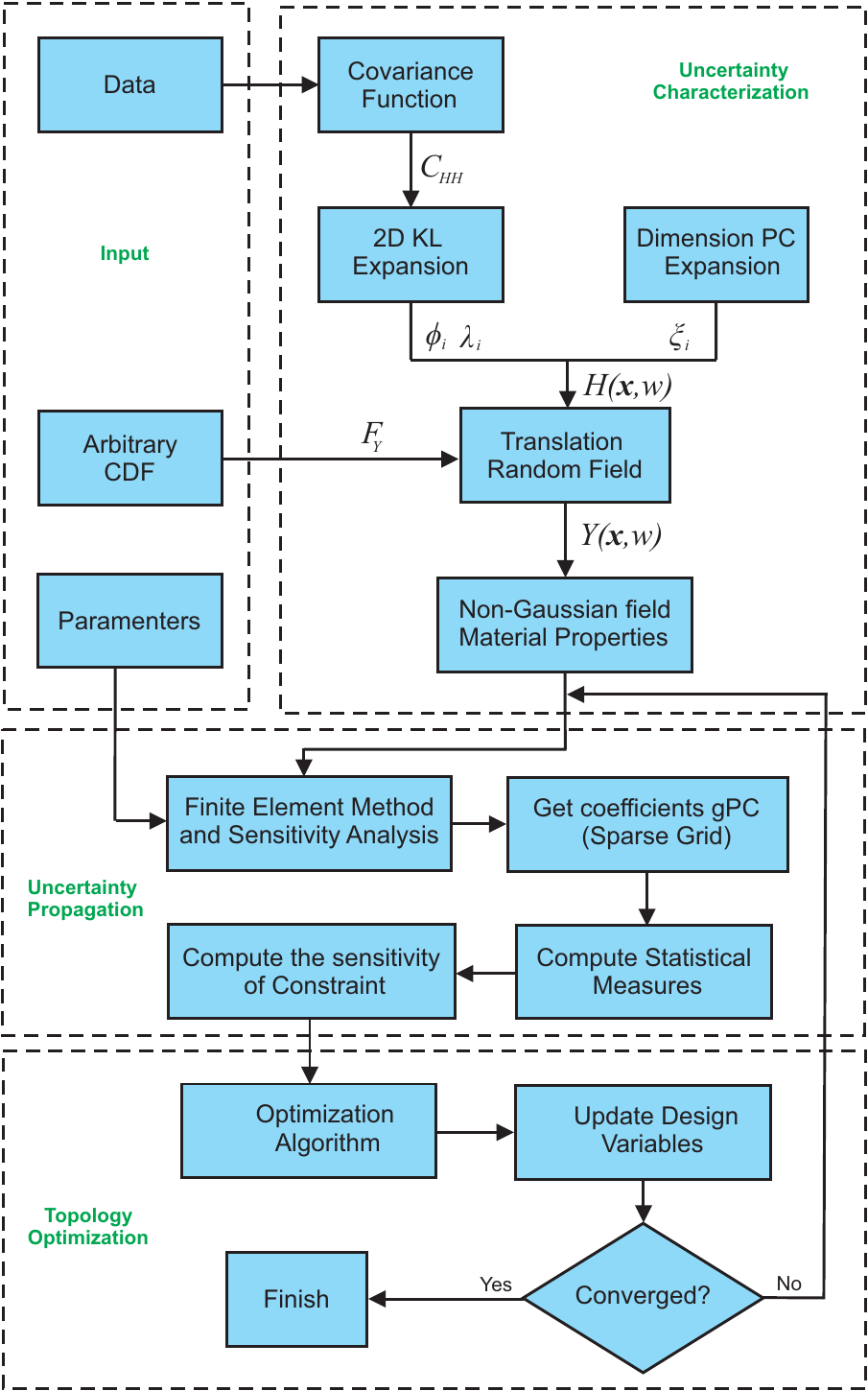}
	\caption{Flowchart of the robust topology optimization algorithm to deal with uncertain material properties.}
	\label{fig04}
\end{figure*}

The topology optimization algorithm for problems with uncertain material properties that is proposed in this work is illustrated as a flowchart  in Figure~\ref{fig04}, and its main steps are described below:

\begin{enumerate}
	\item Initialize the problem to represent the uncertain material properties by using a probabilistic model, i.e. through a non-Gaussian random field;
	\item Use the KL expansion to generate a baseline Gaussian random field from the eigenfunctions and eigenvalues of a given autocovariance function --  Eq.(\ref{15_eq});
	\item Construction of a non-Gaussian field through the memoryless transformation of a given Gaussian field and a desired marginal CDF -- Eq.(\ref{17_eq});
	\item Set the order of the PCE and select the roots of each random variable according to the orthogonal polynomial;
	\item Set the boundary condition for the problem and generate the polygonal element mesh with \texttt{PolyMesher};
	\item Solve the finite element equation $\vec{K}\vec{u}^{(i)} = \vec{f}^{(i)}$ for $ i=1,\ldots,M$
	\item Calculate the sensitivity of the compliance $C$  and with respect to $\rho_e$  according to Eqs.(\ref{46_eq}) and (\ref{51_eq});
	\item Calculate the coefficients $\hat{u}_j$ of the PCE using a sparse grid quadrature;
	\item Calculate the low-order statistical measures according to Eqs.(\ref{46_eq}) and (\ref{47a_eq});
	\item Calculate the sensitivity of the objective function $\tilde{C}$ according to Eqs.(\ref{53_eq}) and (\ref{54_eq});
	\item Update the design variables $\bm{z}$  by the optimizer. Repeat from step 6 until convergence.
\end{enumerate}

The \texttt{PolyTop} framework implemented in MATLAB® \citep{Talischi2012p309} is used for computing the topology optimization. Several codes for polynomial chaos expansion (PCE) and the discretization of random field are developed and integrated to the \texttt{PolyTop}.

\section{Results and discussion}
\label{num_examples}

The effectiveness of the topology optimization framework considering material uncertainties is addressed in this section, where it is applied to 2D numerical structures. Non-Gaussian fields are used to represent the spatial variability of Young’s modulus so that each hexagonal finite element within of domains has its elastic modulus. The robust designs achieved are compared with their deterministic (non-robust) counterpart and with literature results from Tootkaboni et al. \citep{tootkaboni2012topology}, where a Gaussian field is used to represent the material uncertainties.

\subsection{Cantilever beam}

\begin{figure}
	\centering
	\includegraphics[scale=0.75]{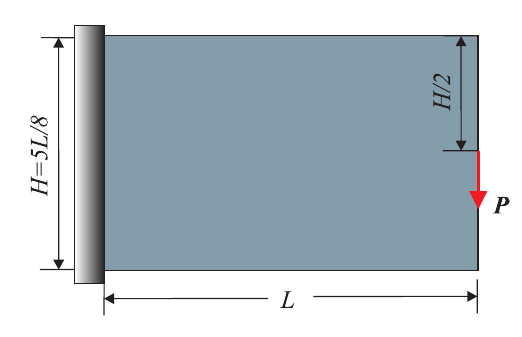}
	\caption{Illustration of the design domain for a 2D cantilever beam.}
	\label{fig05}
\end{figure}

Consider a simple 2D cantilever beam as illustrated in Figure~\ref{fig05}, which is subjected to a vertical point load $P$ applied in the middle right end. The design domain geometry is defined as $L$ long, $5L/8$ height, and $L/40$ thickness. This first example is used to illustrate a comparison between the optimal topologies obtained when uncertainties in material properties are represented by Gaussian and non-Gaussian fields, respectively.

\begin{figure}
	\centering
	\includegraphics[scale=0.75]{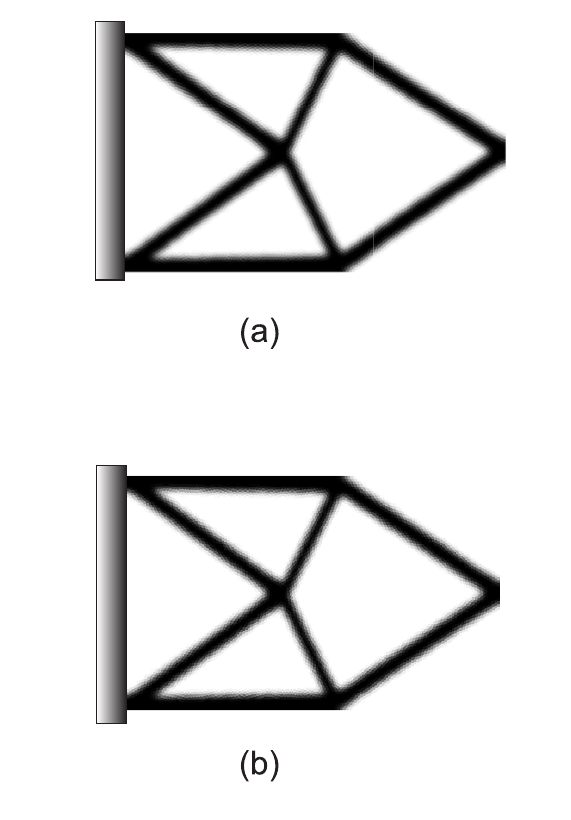}
	\caption{Minimum volume topology for a 2D cantilever beam obtained by: (a) deterministic TO; (b) RTO.}
	\label{fig06}
\end{figure}

The deterministic topology optimization adopts a structure with isotropic material and a constraint of compliance $C = (4000P)/EL$, where $E$ is the Young modulus considered for all elements. The domain is discretized using 11560 polygonal elements, a minimum allowable filter radius of $0.0185L$ is used, and the penalization factor is 3. The values of $P$, $L$, and $E$ are defined as 1, 40, and 100, respectively. Therefore, we obtain the optimal topology shown at the top of Figure~\ref{fig06}, with a minimum volume equal to 0.233.
Because polygonal shape functions are free of the orthogonality constraints that trigger checkerboards on square or triangular grids, the ensuing layouts are obtained \emph{without ad-hoc density filters}, confirming the mesh-stability property.

For robustness, we account for material uncertainty in the Young modulus of each polygonal element. First, we model spatial variability as a 2D Gaussian field with mean $E$ and standard deviation $0.25E$, represented via a KL expansion with correlation lengths $L/2$ and $H/2$, using $N=7$ uncorrelated random variables. The eigenvalues and variance error underlying the KL modes reconstruct approximately 90\% of the field energy. This example is inspired by \citep{tootkaboni2012topology}, where the underlying uncertainty propagation calculation is performed in an intrusive manner. The constraint presented in Eq.(\ref{34_eq}) uses $k=0$, being estimated through PCE of order 3. Thus, we obtain a minimum volume equal to 0.247. The optimal topology of the robust optimization is shown at the bottom of Figure~\ref{fig06}. Although visually close, the two layouts differ by about 6 \% in solid volume (0.233 vs. 0.247) and re-route the main load path through the upper ligament, illustrating that a random field can nudge the optimum away from its deterministic counterpart.

From these results, we see that, for a cantilever beam, the deterministic and robust topologies are very similar when a Gaussian random field represents material uncertainties. But the main idea is to obtain an optimal topology that is more robust to the variability caused by the material uncertainties. Then, using the structure's topology obtained in the deterministic case, we calculate the mean and standard deviation of the volume using ensembles of a Gaussian random field, with values of $0.281$ and $0.181$, respectively.

Table \ref{tab01} shows, for different values of $k$, a comparison between the results obtained from Tootkaboni et al. \citep{tootkaboni2012topology} (first column) and those obtained with our framework (second column). The agreement ($\leq$ 3 \% in both $\mu_{C}$ and $\sigma_{C}$ for every $k$)) confirms that the \emph{non-intrusive} PCE surrogate achieves intrusive-level accuracy while leaving the polygonal FE solver untouched.

\begin{table}
	\centering
	\caption{Comparison of intrusive and non-intrusive methods using log-normal Gaussian field.}
	\vspace{1mm}
	\begin{tabular}{ccccccc}
		\toprule
		&&\multicolumn{2}{c}{Tootkaboni et al. \citep{tootkaboni2012topology}}  && \multicolumn{2}{c}{RTO framework}\\
		\cmidrule(r){3-4} \cmidrule(r){6-7} 
		$k$ && $\mu_{C}$ & $\sigma_{C}$&& $\mu_{C}$ & $\sigma_{C}$\\ 
		\cmidrule(r){1-7} 
		0 &&1.000 &0.131 && 1.000 &0.135\\
		1 &&0.885 &0.115 && 0.881 &0.118\\
		2 &&0.793 &0.103 && 0.787 &0.106\\
		3 &&0.719 &0.094 && 0.712 &0.096\\
		4 &&0.658 &0.086 && 0.649 &0.087\\
		5 &&0.606 &0.079 && 0.597 &0.081\\
		\bottomrule
	\end{tabular}
	\label{tab01}
\end{table}

\begin{figure*}
	\centering
	\includegraphics[scale=0.5]{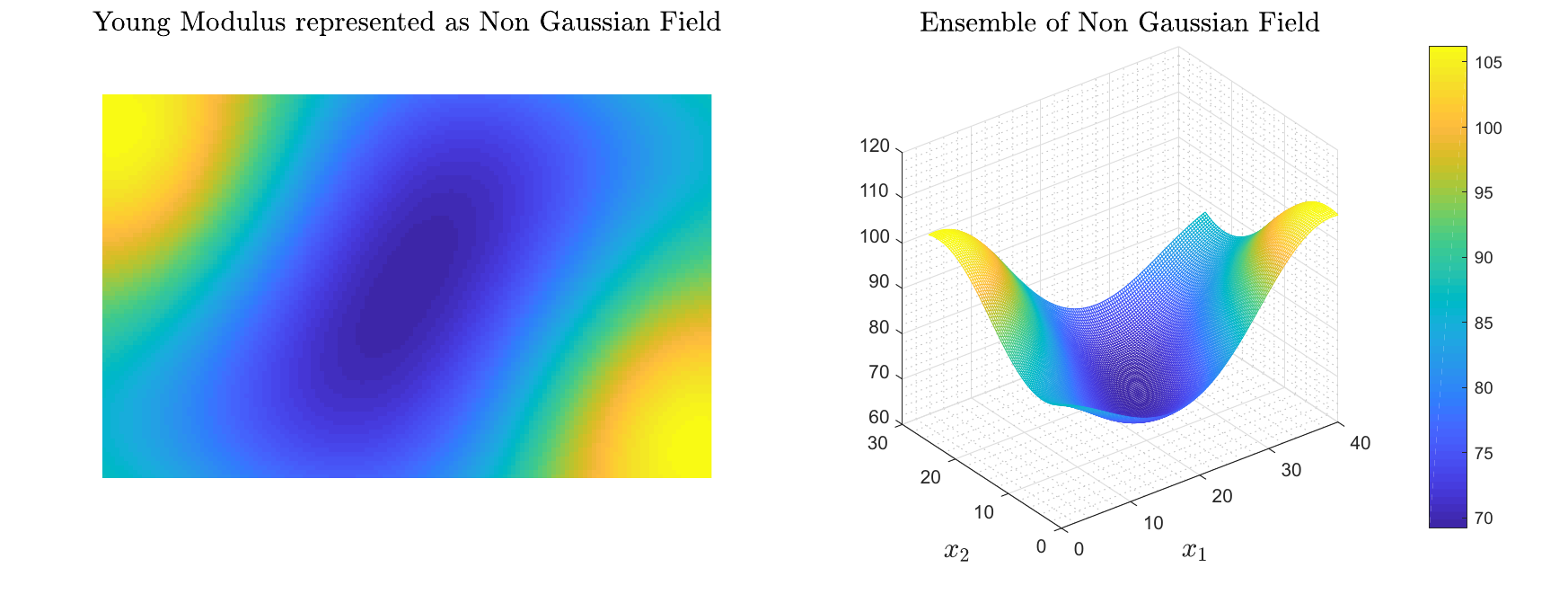}
	\caption{Ensemble of a 2D non-Gaussian field representing the uncertain Young modulus.}
	\label{fig07}
\end{figure*}

\begin{figure*}
	\centering
	\includegraphics[scale=0.82]{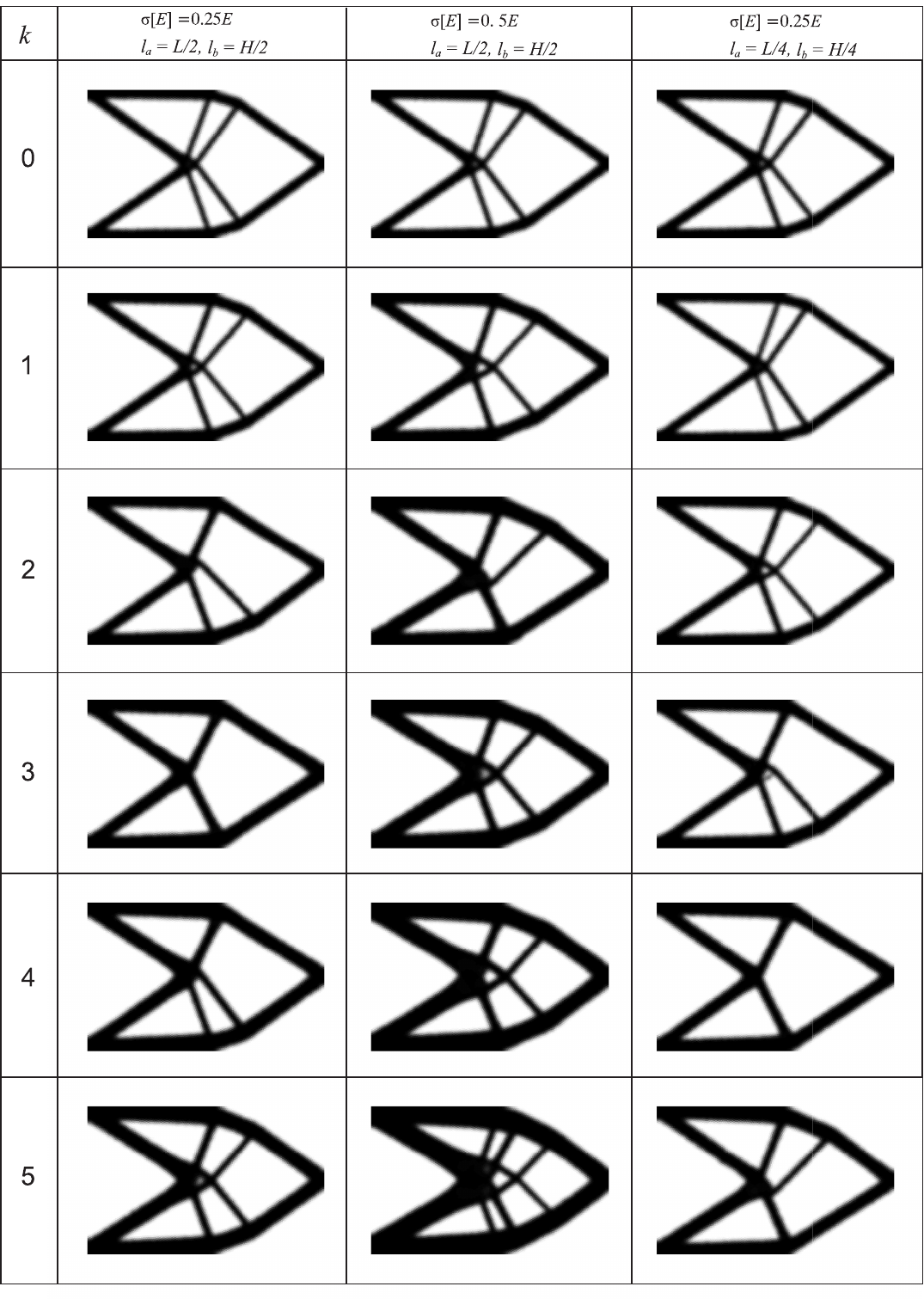}
	\caption{Robust design for a 2D cantilever beam problem, for several values of weighting factor $k$, and different configurations of correlation lengths and standard deviation.}
	\label{fig08}
\end{figure*}

Next, a \emph{strictly positive} non-Gaussian field is considered, ensuring physical consistency of the Young modulus that a Gaussian model cannot guarantee. It is obtained from a Gaussian random field with the autocovariance function given in Eq. ~ (\ref{39_eq}) via a nonlinear transformation using a marginal uniform distribution, as shown in Figure~\ref{fig07}. To visualize the influence of spatial variability in a non-Gaussian field, we increase the standard deviation to $0.35E$ while keeping the same correlation length. In Figure~\ref{fig08}, the first two columns show the robust design for different values of $k$. It may be noted that as the value of $k$ increases, the design becomes more robust and consequently the volume as well. The values obtained are shown in Table~\ref{tab02}. As expected, higher material scatter ($\sigma_{E^0}=0.35E$) or shorter correlation length drives the optimiser to allocate extra material (up to 0.494 vs. 0.249) to secure compliance targets, a trend that is easily captured thanks to the checkerboard-free polygonal mesh. Variation in the correlation length makes the random field more or less correlated and influences discretization. We set the correlation lengths to $l_a=L/4$ and $l_b=H/4$, and the robust design and variation of the results are shown in the third column of Figure~\ref{fig08} and Table~\ref{tab02}, respectively.

\begin{table*}
	\centering
	\caption{Mean value and standard deviation of the compliance, for a 2D cantilever beam, with uncertain Young modulus described as a non-Gaussian random field.}
	\vspace{1mm}
	\begin{tabular}{ccccccccccccc}
		\toprule
		&&\multicolumn{3}{c}{$\sigma_{E^0} = 0.25E$}  && \multicolumn{3}{c}{$\sigma_{E^0} = 0.35E$} && \multicolumn{3}{c}{$\sigma_{E^0} = 0.25E$}\\
		&&\multicolumn{3}{c}{$l_a=L/2$, $l_b=H/2$}  && \multicolumn{3}{c}{$l_a=L/2$, $l_b=H/2$} && \multicolumn{3}{c}{$l_a=L/4$, $l_b=H/4$}\\
		\cmidrule(r){3-5} \cmidrule(r){7-9} \cmidrule(r){11-13}
		$k$ && $\mu_C$ & $\sigma_C$& $V$ && $\mu_C$ & $\sigma_C$& $V$ && $\mu_C$ & $\sigma_C$& $V$\\
		\cmidrule(r){1-13} 
		0 &&1.000 &0.148 &0.249&& 1.000 &0.229&0.259 && 1.000 &0.105 &0.247\\	
		1 &&0.870 &0.129 &0.276&& 0.813 &0.186&0.304 && 0.904 &0.095 &0.265\\
		2 &&0.771 &0.114 &0.302&& 0.684 &0.157&0.349 && 0.826 &0.086 &0.284\\
		3 &&0.691 &0.102 &0.324&& 0.591 &0.136&0.393 && 0.762 &0.079 &0.300\\
		4 &&0.627 &0.093 &0.354&& 0.521 &0.119&0.441 && 0.704 &0.073 &0.315\\
		5 &&0.574 &0.085 &0.385&& 0.466 &0.106&0.494 && 0.656 &0.068 &0.339\\	
		\bottomrule
	\end{tabular}
	\label{tab02}
\end{table*}

It is worth placing these observations in the context of existing results on topology optimization under spatial material uncertainty. Previous studies have reported that introducing random fields in elastic properties may promote the emergence of finer-scale structural members as a means to mitigate local fluctuations in stiffness \citep{silva2016p192}. In the present study, however, the dominant effect of material uncertainty manifests differently. When non-Gaussian random fields with prescribed marginal distributions and correlation structures are employed, the optimized layouts primarily respond through global load-path redistribution and systematic volume reallocation, rather than through a consistent refinement of structural features. This behavior is particularly evident when comparing Gaussian and non-Gaussian representations of the Young’s modulus, where changes in skewness and boundedness of the material distribution induce topological adjustments that cannot be attributed to variance effects alone. These results indicate that higher-order statistical characteristics of the material field play a central role in shaping robust optimal topologies, providing insights that extend beyond those obtained under Gaussian assumptions.

\subsection{Curved beam}

\begin{figure*}
	\centering
	\includegraphics[scale=0.85]{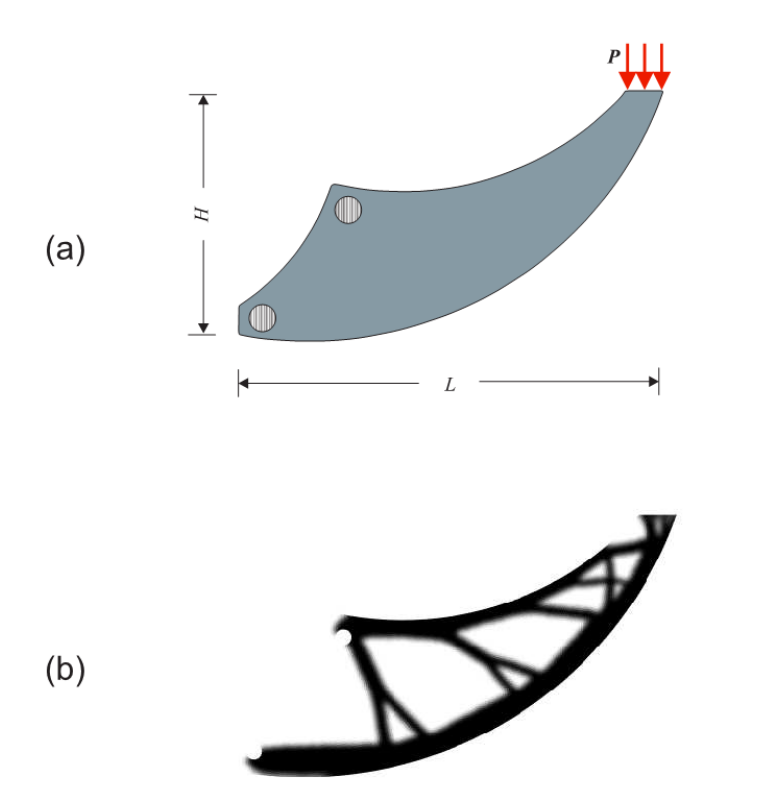}
	\caption{Illustration of a 2D curved beam. (a) Design domain geometry. (b) Minimum compliance topology obtained
by the deterministic TO.}
	\label{fig09}
\end{figure*}

Now consider a simple 2D curved beam subjected to a vertical distributed load $P=0.1u$ applied at the upper corner. The design domain geometry is defined as $L=282u$ long, $H=167.5u$ height, and $L/40$ thickness as illustrated in Figure~\ref{fig09}(a). 

The topology optimization of the curved beam for a deterministic configuration adopts a structure with isotropic material and a constraint of fraction volume $V = 0.45 $, and the Young modulus considered for all elements is $E=100$. The domain is discretized with $10000$ and $40000$ polygonal elements, a minimum allowable filter radius of $5$, and a penalization factor of 3. For the stochastic case, three types of non-Gaussian random fields (Uniform, Gamma, and Beta marginal distributions) are used to represent uncertainties in material properties. The effectiveness of our algorithm is tested by scaling the number of mesh elements to minimize compliance. Therefore, we show in Figure~\ref{fig09}(b), the result of the topology optimization for the deterministic case. 

\begin{figure}
	\centering
	\includegraphics[scale=0.95]{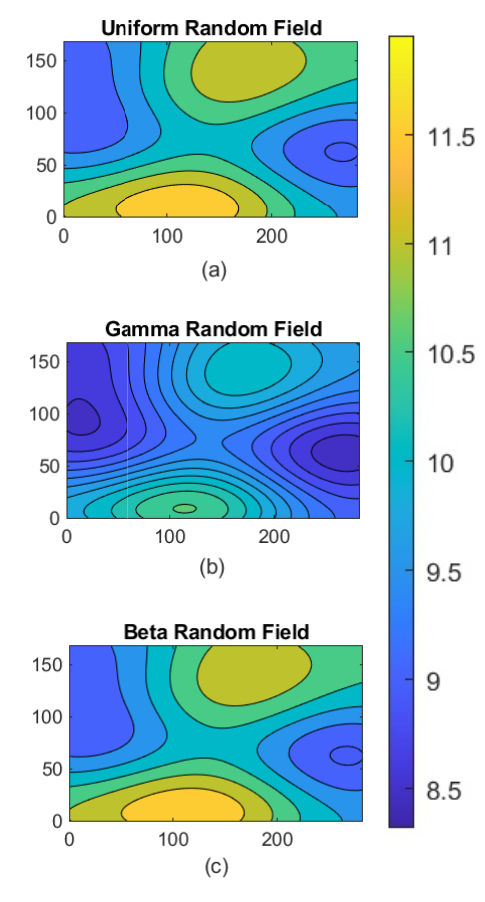}
	\caption{Young modulus represented by a non-Gaussian random field, with three different marginal distributions: (a) Uniform; (b) Gamma; and (c) Beta.}
	\label{fig10}
\end{figure}

Using Eq.(\ref{17_eq}), we develop the translation transformation for obtaining the Uniform, Gamma, and Beta non-Gaussian fields. Those random fields represent uncertainties in material properties. In addition, it is necessary to use a regular geometry that covers the entire domain of the curved structure, thereby facilitating the generation of the random field mesh. Figure~\ref{fig10} shows one realization of the translation random field, using a discrete domain with $10000$ polygonal elements, for each one of the marginal distributions described above.

The realizations shown in Fig.~\ref{fig10} illustrate how different marginal distributions affect the spatial modulation induced by the translation process, while sharing the same underlying Gaussian autocovariance structure. Among the three cases, the Gamma field exhibits the most pronounced variability, which can be attributed to the strong nonlinearity and skewness of its cumulative distribution function. This nonlinear mapping amplifies fluctuations in the upper tail, thereby increasing local contrast.

The Uniform and Beta fields appear visually similar for the parameter choices considered here. This similarity is expected, as both distributions have bounded support and relatively mild nonlinear cumulative distribution functions, which limit extreme realizations and produce smoother spatial modulations of the underlying Gaussian field. The Beta distribution nevertheless allows additional flexibility through its shape parameters, which may 
lead to distinct variability patterns for other parameter settings, even though such differences are not strongly manifested in the present realization.

In all cases, the memory-less translation preserves the spatial correlation structure inherited from the Gaussian field while ensuring the physical admissibility of the material property, namely the positivity of Young’s modulus. These differences and similarities in marginal behavior help explain the variations observed in the robust topology optimization results presented subsequently.

From Fig.~\ref{fig10}(b), it can be seen that the uncertainty variability for the Gamma field is more visible than the other fields. This characteristic is because of the high nonlinearity of the prescribed marginal CDF. In all three cases, the polygonal discretization maintains clean, artefact-free contours; no post-processing filters were necessary.

\begin{figure}
	\centering
	\includegraphics[scale=0.9]{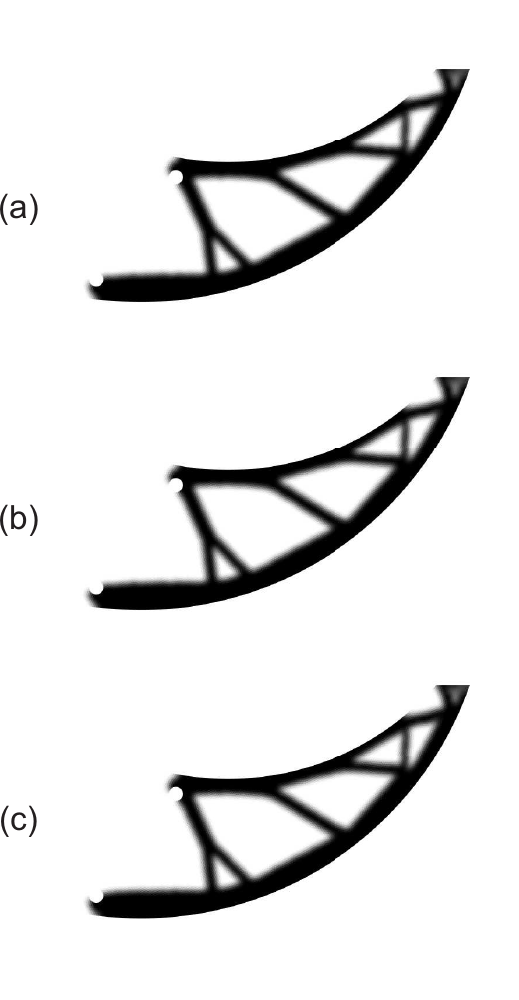}
	\caption{Robust topologies for a 2D curved beam with the Young modulus represented by a non-Gaussian random field, with three different marginal distributions: (a) Uniform; (b) Gamma; and (c) Beta.}
	\label{fig11}
\end{figure}

Figure~\ref{fig11} shows the results using the three non-Gaussian random fields to minimize the compliance, and Table~\ref{tab03} shows the results obtained for each value of $k$. It can be seen that the variability decreases as $k$ rises. Moreover, the choice of marginal distribution matters: for $k=0$ the Beta field yields a 32 \% lower mean compliance (7.20 vs. 10.56) and a 72 \% lower standard deviation compared with the Uniform field, underscoring the design sensitivity to non-Gaussian statistics.

\begin{table*}
	\centering
	\caption{Mean value and standard deviation of the compliance, for a 2D curved beam, with uncertain Young modulus described as a non-Gaussian random field, for different marginal distributions.}
	\vspace{1mm}
	\begin{tabular}{cccccccccc}
		\toprule
		&&\multicolumn{2}{c}{$\mbox{Uniform}$}  && \multicolumn{2}{c}{$\mbox{Gamma}$} && \multicolumn{2}{c}{$\mbox{Beta}$}\\
		
		\cmidrule(r){3-4} \cmidrule(r){6-7} \cmidrule(r){9-10}
		$k$ && $\mu_C$ & $\sigma_C$&& $\mu_C$ & $\sigma_C$ && $\mu_C$ & $\sigma_C$\\
		\cmidrule(r){1-10} 
		0 &&10.5562 &2.3608 && 9.6289 &2.1441 &&7.1956 & 0.6534 \\	
		1 &&10.5595 &2.3534 && 9.6285 &2.1379 && 7.1950 &0.6526 \\
		2 &&10.5650 &2.3492 &&9.6332 &2.1348 &&7.1958 &0.6518 \\
		\bottomrule
	\end{tabular}
	\label{tab03}
\end{table*}

It is worth emphasizing that the present framework adopts a density-based topology optimization formulation by design. The objective of this work is not to enforce strictly binary (0--1) layouts, but to assess how spatially correlated, non-Gaussian material uncertainty influences optimal load paths, volume allocation, and compliance statistics. In this context, the occurrence of intermediate-density regions in the reported layouts is a direct consequence of the continuous density parametrization adopted in this study and of the robust optimization setting in which statistical performance measures are evaluated. Such intermediate values are not intended to represent physical material phases, but rather arise from the 
relaxed design space commonly employed in density-based formulations to ensure well-posed sensitivity analysis and stable convergence under uncertainty. Accordingly, the presence of these regions does not affect the interpretation of the results, which focus on load redistribution, volume allocation, and compliance statistics under non-Gaussian material variability.

Alternative formulations, such as boundary-based or level-set approaches, enforce sharp material interfaces by construction and constitute a complementary modeling choice. The extension of the present framework to such formulations, combined with non-Gaussian random fields defined on polygonal discretizations, represents a promising avenue for future research when strict phase separation is required.

Unlike Gaussian-based robust formulations, the present results demonstrate that non-Gaussian material statistics can modify optimal designs through mechanisms that are not reducible to local feature refinement, underscoring the importance of modeling higher-order uncertainty effects.

\section{Conclusions}
\label{concl_remaks}

This work advances robust structural design by fusing three complementary ingredients—polygonal finite elements, a physically consistent non-Gaussian representation of material variability, and a non-intrusive polynomial-chaos surrogate—within a single topology-optimization loop. The polygonal mesh mitigates checkerboard artefacts and mesh bias in a stable manner, the non-Gaussian random field enforces the positivity of Young’s modulus and captures heavy-tailed or bounded statistics, and the sparse PCE delivers compliance moments at a fraction of the cost of full stochastic analyses.

\begin{enumerate}
  \item \emph{Efficiency without intrusiveness.}  
  The surrogate reproduces intrusive reference solutions to within a few percent while leaving the deterministic solver unmodified, keeping the wall-clock effort of the robust loop comparable to that of a handful of deterministic re-analyses.

  \item \emph{When Gaussian $\approx$ deterministic.}  
  If material variability is idealised as Gaussian, the optimal layout changes little and the deterministic design may already be near-robust—confirming earlier studies and providing a useful sanity check for the framework.

  \item \emph{When non-Gaussian $\neq$ Gaussian.}  
  Modelling the modulus as a Uniform, Gamma, or Beta field triggers a visible reallocation of 6–12\,\% of the solid volume, alters load paths, and reduces compliance scatter.  
  These differences underline that an accurate, physically admissible uncertainty model is essential for genuinely robust structures. Unlike Gaussian-based robust formulations, these changes arise primarily from load-path redistribution and volume reallocation rather than systematic feature refinement, highlighting the role of higher-order material statistics.

    \item \emph{Robustness under discretization changes.}  
    The robust layouts remain stable under mesh refinement and unstructured discretizations, indicating that the polygonal formulation and the random-field mapping interact favourably in the presence of spatial material uncertainty.
\end{enumerate}

\paragraph{Limitations and future work.}
The present study focused on two-dimensional benchmarks; extending the framework to large-scale three-dimensional problems and manufacturing-oriented constraints (e.g., overhang angles, print-path continuity) is an immediate next step.  
In addition, integrating model-form uncertainty—for instance via stochastic reduced bases or Bayesian material calibration—would further strengthen the link between simulation and fabricated parts.

\medskip
Overall, the proposed combination of polygonal meshes, non-Gaussian random fields, and non-intrusive PCE provides a practical, computationally affordable route to topology-optimised designs that remain reliable under realistic material uncertainty.

\section*{Funding}

This research received financial support from the Tecgraf/PUC-Rio, the National Council for Scientific and Technological Development (CNPq) grants 317319/2021-3 and 305476/2022-0; the Coordination for the Improvement of Higher Education Personnel (CAPES), Finance Code 001; and the Carlos Chagas Filho Research Foundation of Rio de Janeiro State (FAPERJ) under grants 211.037/2019, and 204.477/2024.

\section*{Compliance with ethical standards}

\section*{Conflict of Interest }
The authors declare that they have no conflict of interest.

\bibliographystyle{spbasic} 

\bibliography{References} 

\end{document}